\documentclass{PoS}

\usepackage{graphicx}
\graphicspath{{figs/}}
\newcommand{\ket}[1]{\left| #1 \right\rangle}
\newcommand{\bra}[1]{\left\langle #1 \right|}

\newcommand{\comma}{\ ,}
\newcommand{\point}{\ .}

\newcommand{\Mst}{a\sqrt{\sigma}}

\title{
\vspace*{-5cm}
{\normalsize {\rm \hfill{ Edinburgh 2013/23} }} \\
\vspace*{5cm}
Gluonic observables and the scalar spectrum of \\ twelve--flavour QCD}

\ShortTitle{Gluonic observables and the scalar spectrum of twelve--flavour QCD}

\author{
  Yasumichi Aoki$^a$,
  Tatsumi Aoyama$^a$,
  Masafumi Kurachi$^a$,
  Toshihide Maskawa$^a$,
  Kohtaroh Miura$^a$,
  Kei-ichi Nagai$^a$,
  Hiroshi Ohki$^a$, 
  \speaker{Enrico Rinaldi}$^b$\thanks{E-mail: E.Rinaldi@sms.ed.ac.uk},
  Akihiro Shibata$^c$,
  Koichi Yamawaki$^a$,
  and 
  Takeshi Yamazaki$^a$ 
  
  \hspace*{55mm} (LatKMI Collaboration) 
  \\
  
  $^a$
  Kobayashi-Maskawa Institute for the Origin
  of Particles and the Universe (KMI), Nagoya University, Nagoya
  464-8602, Japan \\
  $^b$ 
  Higgs Centre for Theoretical Physics, SUPA, School of Physics and Astronomy, 
  University of Edinburgh, Edinburgh EH9 3JZ, UK \\
  $^c$
  Computing Research Center, High Energy Accelerator Research Organization (KEK), 
  Tsukuba 305-0801, Japan
}

\abstract{
We measure glueball masses and the string tension in twelve--flavour QCD,
aiming at comparing the emerging gluonic spectrum to the mesonic one.
When approaching the critical surface at zero quark mass, the hierarchy of masses in the different 
sectors of the spectrum gives a new handle to determine the existence of an infrared fixed point.\\
We describe the details of our gluonic measurements and the results obtained on a 
large number of gauge configurations generated with the HISQ action.
In particular, we focus on the scalar glueball and its mixing with a flavour--singlet
fermionic state, which is lighter than the pseudoscalar (would--be pion) state.
The results are interesting in view of a light composite Higgs boson in walking technicolor theories.
}

\FullConference{31st International Symposium on Lattice Field Theory - LATTICE 2013\\
		July 29 - August 3, 2013\\
		Mainz, Germany}

\begin{document}

\section{Introduction}
\label{sect:intro}

Many lattice studies of many--flavour QCD have been carried out in recent years to determine whether the SU(3) gauge theory with massless fermions is chirally broken or conformal at large distances. The importance of the answer to such question is related to the theory being or not a good candidate for a Walking Technicolor (WTC) model, featuring an approximate scale invariance and a mass anomalous dimension $\gamma_m \sim 1$~\cite{Yamawaki:1985zg}. If a WTC candidate can be found with the above requirements and, additionally, with a light flavour--singlet scalar bound state, then it could be able to explain the recent Higgs boson discovery, in a scenario that goes beyond the Standard Model~\cite{Matsuzaki:2012mk}.

For this purpose, the lattice framework proves to be an excellent tool due to the inherent non--perturbative nature of infrared dynamics in non--abelian gauge theories. It is important to stress that lattice studies are affected by systematic effects which need to be estimated, in particular for WTC candidate theories (for details on lattice of WTC models see Ref.~\cite{plenary}.)

In this proceeding we focus on an instance of many--flavour QCD, which has been intensively studied by several groups in the past: $N_f=12$ QCD. A previous study by us~\cite{Aoki:2012eq} has shown QCD with $N_f=12$ massless fermions to be consistent with having an infrared conformal fixed point. At the same time, the spectrum of the theory contains a flavour--singlet scalar particle~\cite{Aoki:2013zsa}, lighter than the would--be pion, when the conformal symmetry is explicitly broken by a fermion mass term $\propto m_f \bar{\psi}\psi$. Such a light scalar state could be due to the claimed original scale invariance of the theory and would become a techni--dilaton behaving as a Higgs--like particle if conformal symmetry was spontaneously broken by a dynamical fermion mass, which would play a role dynamically similar to the explicit fermion mass in the conformal phase. The latter option seems to hold for the $N_f=8$ QCD theory~\cite{Aoki:2013xza,Aoki:2013ttt}, where chiral symmetry breaking and a light flavour--singlet scalar bound state appear together as signs of a possible WTC candidate.

In the following, we look at large--distance features of the twelve--flavour theory which have not been particularly investigated so far. Namely, we measure gluonic observables, like the spectrum of scalar glueballs and the string tension. Although there exists a study of the string tension for this theory~\cite{Fodor:2011tu}, we add an independent measurement and we also compare our findings to the fermionic part of the spectrum in the scalar, pseudoscalar and vector channels. 
The results presented in the following must be considered as preliminary.

\subsection{Lattice setup}
\label{sec:lattice-setup}

We introduce the details of our lattice simulations, which were also used to obtain the results of Ref.~\cite{Aoki:2013zsa}. The numerical simulations of the continuum SU(3) gauge theory with 12 degenerate fermions are carried over by the LatKMI collaboration using the Highly Improved Staggered Quark (HISQ) action and the tree--level Symanzik gauge action. The $N_f=12$ degenerate massless fermions in the continuum are represented using $N_s=3$ degenerate staggered fermion species of mass $am_f$, each coming in $N_t=4$ tastes ($N_f=N_s \times N_t$) on the lattice, where $a$ is the lattice spacing.

The bare gauge coupling constant is set to $\beta=6/g^2=4.0$, which is the value closer to weak coupling in Ref.~\cite{Aoki:2012eq}. Hence, the results presented in this proceeding are obtained at fixed lattice spacing. We simulate four physical volumes $(aL)^3$ with $L=18, \, 24, \, 30, \, 36$ and aspect ratio $T/L = 4/3$, where $T$ is the number of points in the temporal direction, and we select different bare quark masses on each volume, from $am_f = 0.05$ on the largest $L=36$ volume, to $am_f=0.16$ on the smallest $L=18$. Using a larger volume at smaller bare fermion mass is important to keep finite--size effects under control, as we describe in the following.

This set of parameters allows us to check for finite--size systematics at fixed quark mass and to test the fermion mass dependence of the spectrum in the infinite volume limit. As a reference, the simulated pseudoscalar (would--be pion) masses are $0.32 < am_\pi < 0.74$, while the decay constant reads $0.06 < aF_\pi < 0.14$.

The most important features of our simulations, that is worth mentioning, are the good flavour symmetry realisation of the HISQ action and the large number of configurations, obtained after $\mathcal{O}(1000)$ thermalisation trajectories, from uninterrupted Markov chains. In fact, we generate between approximately $10000$ and $30000$ trajectories to perform a large statistics analysis of the gluonic spectrum.

\section{Gluonic observables}
\label{sect:glue}

In this section we briefly introduce our methodology to extract the mass of a flavour--singlet scalar bound state with $0^{++}$ quantum numbers using gluonic operators. We employ correlators of spatial Wilson loops symmetrised to have the desired scalar quantum numbers. We refer to the state extracted this way as the scalar glueball because such operators create a gluonic bound state from the vacuum of a pure Yang--Mills theory on the lattice. In twelve--flavour QCD, where dynamical fermions are affecting the dynamics of the vacuum, there could be a bound state with the same quantum numbers but fermionic content $\approx \bar{q}q$, that can in principle have a non--zero overlap on the gluonic operators. We have measured such a $\bar{q}q$ state using fermionic bilinear operators with the desired quantum numbers in Ref.~\cite{Aoki:2013zsa}.

We use a large number of different shapes for the spatial Wilson loops in order to enlarge the variational operatorial basis and we also include operators with different levels of smearing and blocking. This way we obtain a large variational basis of gauge--invariant interpolating operators $\mathcal{O}_{\alpha}(x,t)$ with well--defined rotational quantum number.

For every configuration, we measure a matrix of correlation functions, whose elements are
\begin{equation}
  \label{eq:corr-matrix}
  \tilde{C}_{\alpha\beta}(t) \; = \; \sum_\tau \bra{0} \mathcal{O}_{\alpha}^{\dag} (t+\tau) \mathcal{O}_{\beta}(\tau) \ket{0} \point
\end{equation}
By solving the generalised eigenvalue problem for the matrix above, optimal operators (i.e. those that create almost pure states $\ket{i}$) can be found that are a linear sum of the basis vectors
\begin{equation}
  \label{eq:optimal-op}
  \tilde{\mathcal{O}}_{i} (t) \; = \; \sum_\alpha v^i_\alpha \mathcal{O}_\alpha(t)
  \, ; \qquad
  \tilde{\mathcal{O}}_{i} (t)\ket{0} \approx \ket{i}
  \comma
\end{equation}
where $v^i_\alpha$ are the components of the $i^{\rm th}$ eigenvector. Different eigenvectors $v^i$ correspond to different states.

The mass $am_i$ of the $i^{\rm th}$ state is extracted by fitting correlators of optimal operators using
\begin{equation}
  \label{eq:coshFIT}
  \bar{C}_{ii}(t) = |c_i|^2 \left( e^{-am_{i}t} + e^{-am_{i}(T-t)}\right) \comma
\end{equation}
where $T$ is the length of the lattice in the time direction and the functional form is a consequence of the usual exponential decay in a lattice with periodic boundary condition in the time direction. In the next section, we refer to the mass extracted from the above fitting function as $aM_G$. The procedure described above is quite standard in lattice spectroscopy and more technical details can be found in Ref.~\cite{Lucini:2010nv}.

By changing the operators in the correlation function, one can study different states. In particular, a useful channel we investigate with this variational technique is the string tension in units of the lattice spacing, $\Mst$. One way of extracting this observable requires to estimate the ground--state mass of torelons. Torelons are finite--volume excitations which wrap around the periodic boundaries of the lattice; they couple to spatial Polyakov loop operators, which we smear to create a variational basis.

The torelon mass relates to the string tension through
\begin{equation}
  \label{eq:sigma}
  am_{\rm tor}(L)  =  a^2\sigma L - \frac{\pi}{3L} - \frac{\pi^2}{18L^3}\frac{1}{a^2\sigma} \comma
\end{equation}
where both the leading order term and the next--to--leading order universal correction~\cite{string} are included. Once the torelon mass has been estimated for a fixed lattice size $L$, Eq.~(\ref{eq:sigma}) is inverted to obtain $\Mst$. In the next section we show how well we are able to extract $am_{\rm tor}(L)$ from spatial Polyakov loop correlators and what is the size of finite--volume effects on the string tension.

\section{Results}
\label{sect:results}

\begin{figure}[!h]
  \centering
  \includegraphics[width=0.45\textwidth]{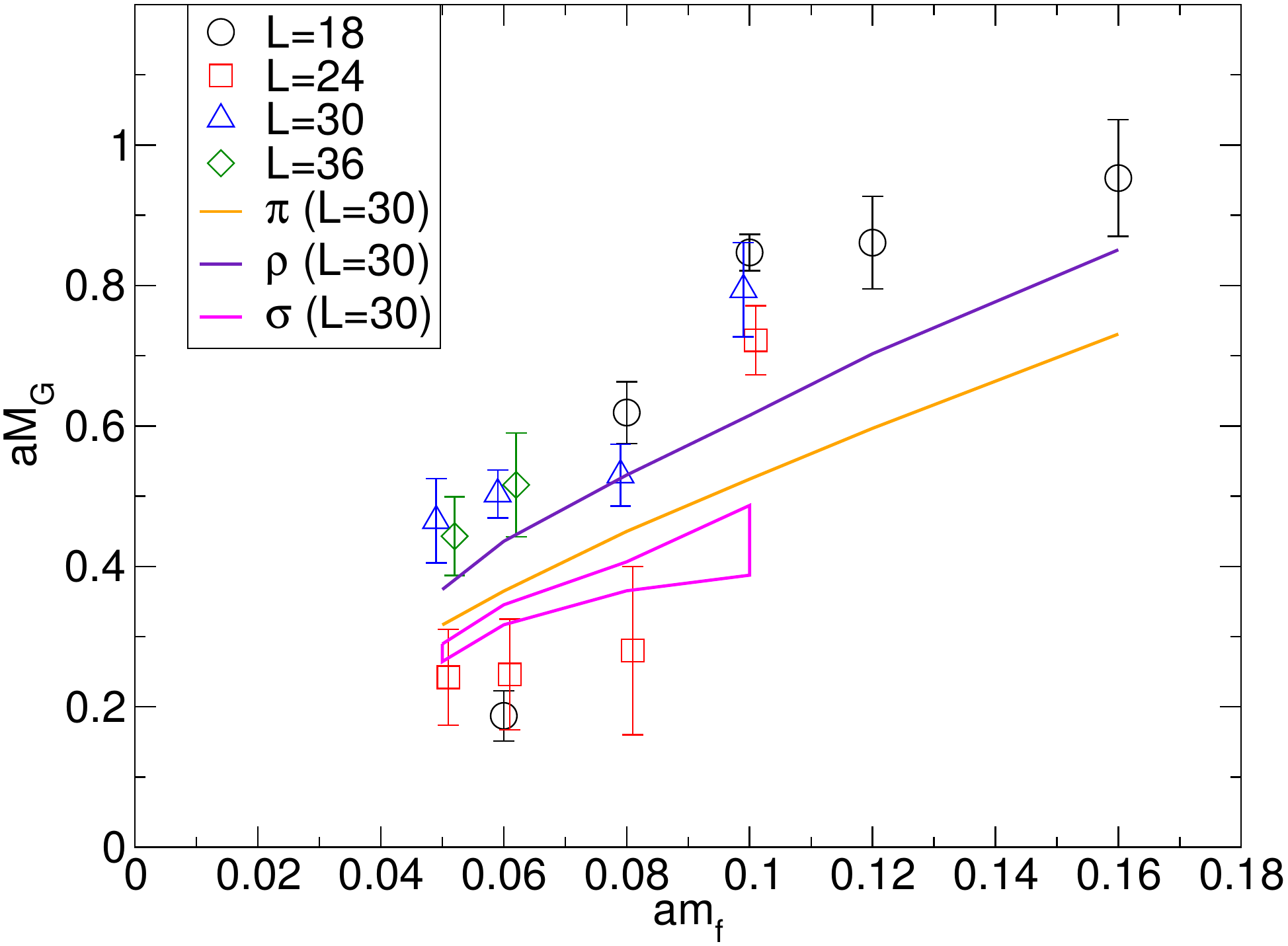}
  \caption{The scalar glueball mass in units of the lattice spacing $aM_G$ is plotted as a function of the bare quark mass $am_f$ for several volumes (horizontally shifted for clarity). The pseudoscalar ($\pi$) and vector ($\rho$) state masses are highlighted with solid lines (from Ref.~\cite{Aoki:2012eq}). The area delimited by a magenta line shows the location of the scalar ($\sigma$) state (from Ref.~\cite{Aoki:2013zsa}).}
  \label{fig:glueball-all-volumes}
\end{figure}
The spectrum of the scalar glueball mass $aM_G$ is obtained for all the investigated parameters and it is plotted in Fig.~\ref{fig:glueball-all-volumes}. At rather large bare fermion masses $am_f > 0.1$, we only have a single volume estimate which shows a heavy glueball mass $aM_G \sim 0.8$. For lighter fermions we are able to check the volume dependence of $aM_G$ and we also observe a clear fermion mass dependence, with the glueball state becoming lighter towards the chiral limit. This last observation indicates that fermion loop contributions are affecting the spectrum, since such a dependence could not be accommodated for in a quenched theory (remember that all results presented here are for a fixed ultraviolet cutoff).

In the light fermion region $am_f < 0.1$ we observe a degeneracy between the scalar glueball mass and the scalar flavour--singlet mass of the $\sigma$ state. This is somehow expected since our gluonic operators in Sec.~\ref{sect:glue} have the same quantum numbers of the fermionic ones used in Ref.~\cite{Aoki:2013zsa}. However, we are able to resolve a light scalar state with gluonic operators only for a limited range of volumes and fermion masses. A better estimate of the coupling between gluonic and fermionic d.o.f. in the $\sigma$ state is possible by including fermion operators in the variational basis of Eq.~(\ref{eq:corr-matrix}): such analysis is in progress and will be reported elsewhere. Nonetheless, our results are a first observation of a flavour--singlet scalar state lighter than the pseudoscalar and vector state using gluonic and fermionic correlators in a (near-)conformal theory. 
\begin{figure}[h]
  \centering
  \begin{tabular}{cc}
    \includegraphics[width=0.45\textwidth]{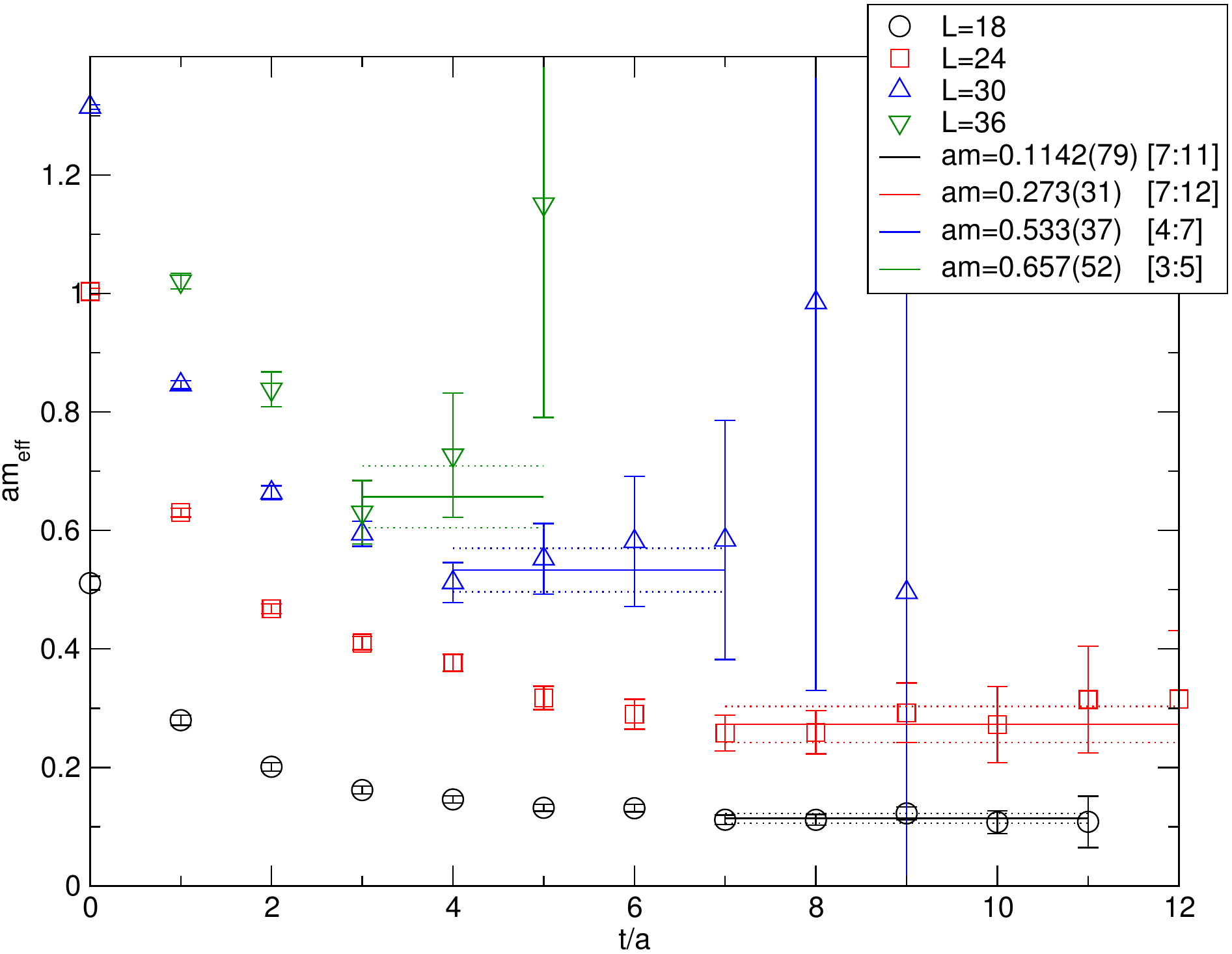} &
    \includegraphics[width=0.45\textwidth]{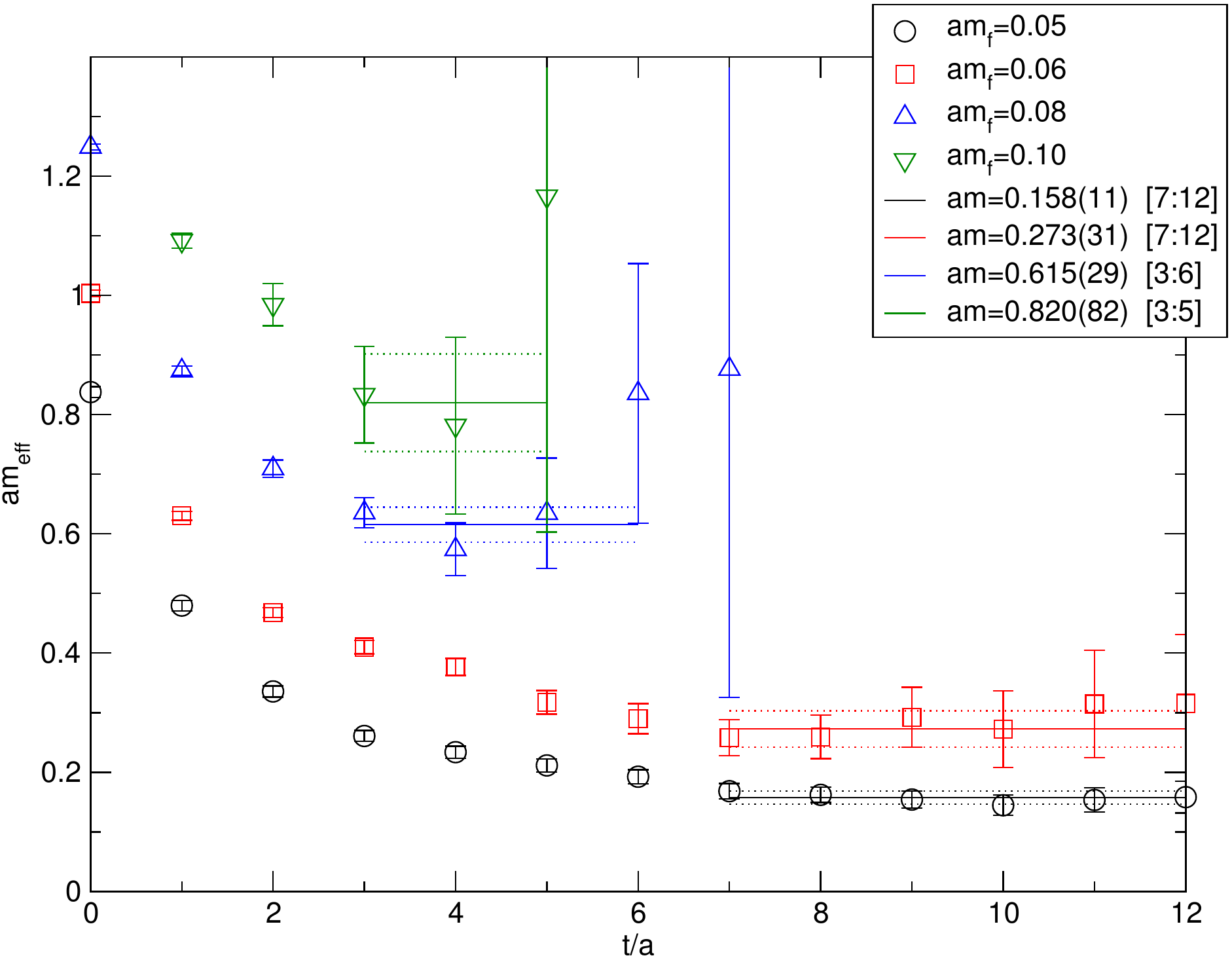} \\
  \end{tabular}
  \caption{(Left) The ground state effective mass extracted from correlators of spatial Polyakov loops with a variational approach. Results for different spatial volumes $L=18$, $24$, $30$ and $36$ are shown at fixed bare quark mass $am_f=0.06$. (Right) Same as the left panel but for different bare quark masses $am_f \leq 0.10$ at fixed lattice volume $L=24$. A good plateaux is always found in both cases and the fitted results are shown with solid lines.}
  \label{fig:torelon-meff-am06}
\end{figure}

We now move to discuss the results for the string tension. In Fig.~\ref{fig:torelon-meff-am06} we show the effective mass coming from spatial Polyakov loop correlators. When a plateaux is reached in the effective mass, the correlator is expected to behave as in Eq.~(\ref{eq:coshFIT}). The fitted mass shown on top of the data points in Fig.~\ref{fig:torelon-meff-am06} is our best estimate for $am_{\rm tor}(L)$. The plots show both the $L$ dependence of the torelon mass at fixed bare fermion mass, resulting in a volume dependence of the extracted string tension, and the mass dependence at fixed volume.

At each bare fermion mass $am_f$ where results on more than one volume are present, we are able to find at least two points compatible within the statistical errors; among them, we use the largest volume result as our infinite--volume estimate for $\Mst$. We observe that the string tension is free from finite--size effects when long Polyakov loops are considered, $L\Mst \gtrsim 3$.
\begin{figure}[!h]
  \centering
  \begin{tabular}{cc}
  \includegraphics[width=0.45\textwidth]{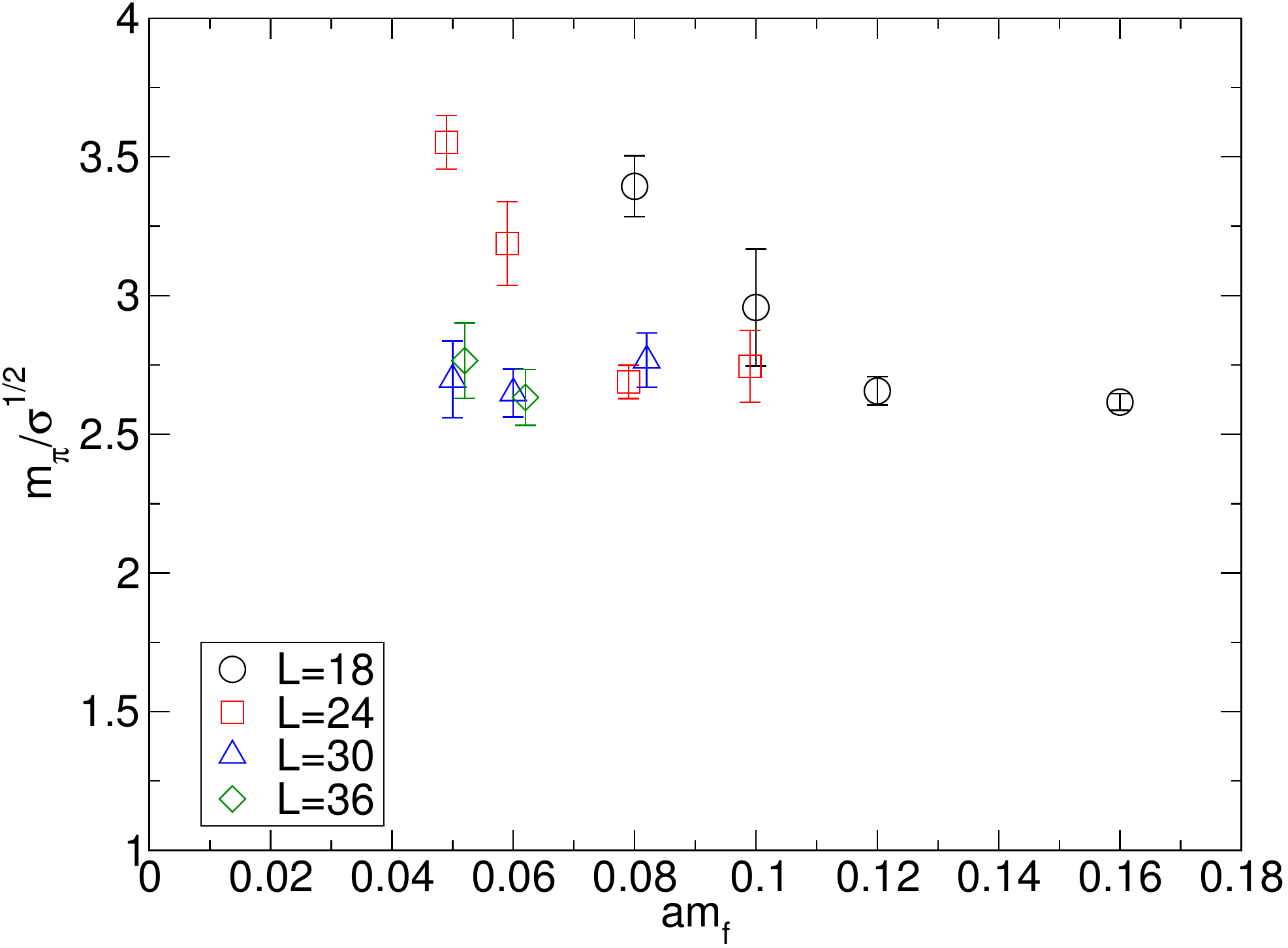} &
  \includegraphics[width=0.45\textwidth]{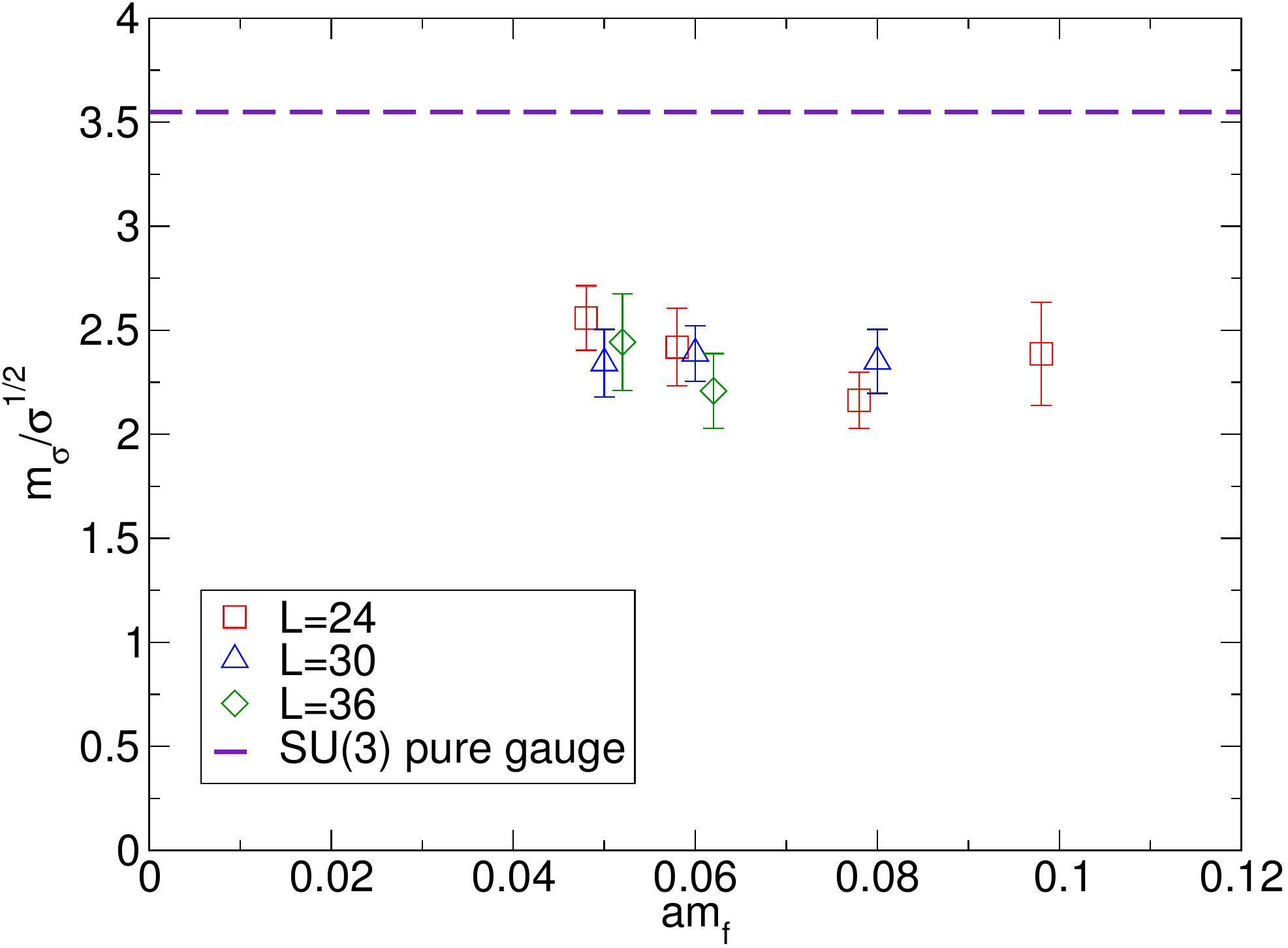} \\
\end{tabular}
\caption{(Left) The mass ratio between the pseudoscalar state and the string tension $m_{\pi}/\sqrt{\sigma}$ for different bare fermion masses. (Right) The mass ratio between the fermionic flavour--singlet scalar state and the string tension $m_{\sigma}/\sqrt{\sigma}$. As a comparison, the ratio between the lightest scalar state and the string tension of a SU(3) pure gauge theory in the continuum is shown. Data on different volumes are shown (slightly displaced for clarity).}
  \label{fig:ratios}
\end{figure}

In order to gain some insights on the (near-)conformal nature of $N_f=12$ QCD, we look at ratios of spectral quantities: in the conformal phase, all masses in the spectrum vanish with the same power law as a function of the fermion mass; therefore, ratios of masses remain constant when approaching the chiral limit of the theory. In Fig.~\ref{fig:ratios} we plot the ratio $m_{\pi}/\sqrt{\sigma}$ between the pseudoscalar mass from Ref.~\cite{Aoki:2012eq} and the string tension, and also the ratio $m_{\sigma}/\sqrt{\sigma}$ between the scalar meson mass from Ref.~\cite{Aoki:2013zsa} and the string tension. Results for different volumes are reported to show that negligible finite--size effects are present when larger volumes are considered towards the chiral limit.

The first ratio we consider, $m_{\pi}/\sqrt{\sigma}$, is well compatible with a constant value all the way to the lightest fermion masses, and the same applies to the second one, $m_{\sigma}/\sqrt{\sigma}$. This result suggests a common scaling of all the studied quantities toward the chiral limit, in contrast with the behaviour from chiral perturbation theory.
\begin{figure}[!h]
  \centering
    \includegraphics[width=0.45\textwidth]{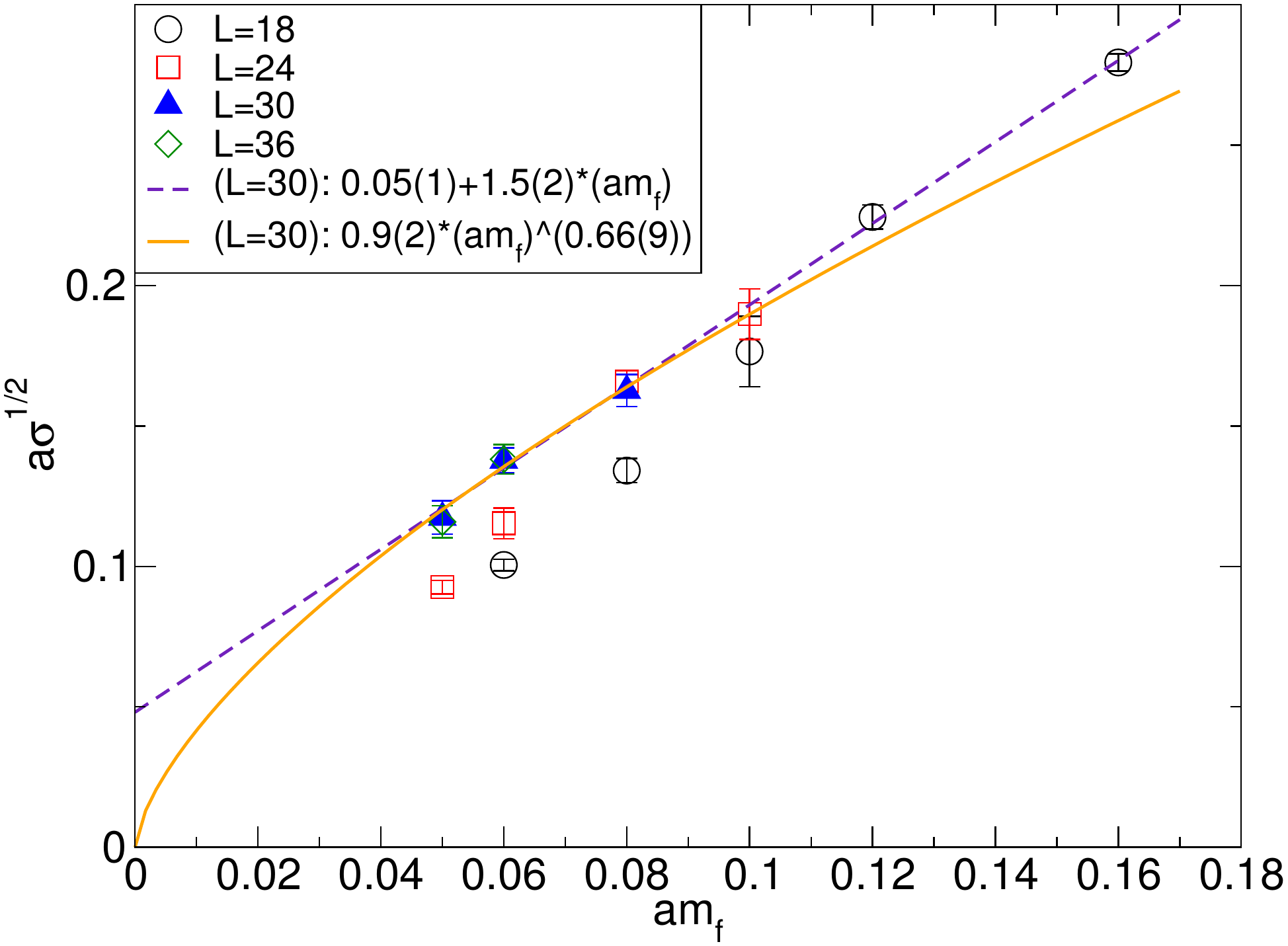}
  \caption{The string tension in units of the lattice spacing $\Mst$ is plotted as a function of the bare quark mass $am_f$. Estimates from different volumes from $L=18$ to $L=36$ are plotted. Two fits of the $L=30$ data (only 3 points) are included: the dashed line is a linear fit implying a non--zero $\sqrt{\sigma}$ in the chiral limit, while the solid curve is a hyperscaling relation with $\gamma \sim 0.5$.}
  \label{fig:string}
\end{figure}

\section{Discussion and conclusions}
\label{sect:concl}

In this proceeding we have presented our preliminary results on large volumes for the scalar glueball mass and the string tension in twelve--flavour QCD. The scalar glueball obtained from a variational analysis including gluonic operators with scalar quantum numbers appears to be heavier than the pseudoscalar and vector mesons when larger volumes are considered towards the chiral limit. However, on smaller volumes and at light fermion masses, the extracted glueball mass is degenerate with the one independently obtained using fermionic interpolating operators only; the latter has been interpreted as a light flavour--singlet mesonic state~\cite{Aoki:2013zsa}.

We also measured the string tension using correlator of smeared spatial Polyakov loops. We then considered mass ratios where the string tension was used to set the scale. Both the pseudoscalar and the scalar state in units of the string tension show a constant behaviour in the explored mass range and large volume limit. This indicates a regime where all energy scales in the spectrum have the same functional dependence on the bare fermion mass. This is consistent with the theory being infrared conformal in the chiral limit, as expected from a previous study of the mesonic spectrum alone~\cite{Aoki:2012eq}.

By considering the string tension alone, as it is shown in Fig.~\ref{fig:string}, a definite conclusion on the nature of the massless theory could not be reached. In fact, both the conformal hypothesis featuring hyperscaling, and the confining hypothesis with a finite mass gap in the chiral limit are compatible with the lattice data (see caption for details on the fits).

\vspace{0.5cm}
{\it Acknowledgments--} Numerical calculations have been carried out on the high--performance computing system $\varphi$ at KMI, Nagoya University,  and on the computer facilities of the Research Institute for Information Technology in Kyushu University. This work is supported by the JSPS Grant-in-Aid for Scientific Research  (S) No.22224003, (C) No.23540300 (K.Y.), for Young Scientists (B) No.25800139 (H.O.) and No.25800138 (T.Y.), and also by Grants-in-Aid of the Japanese Ministry for Scientific Research on Innovative Areas No.23105708 (T.Y.). E.R. was supported by a SUPA Prize Studentship and a FY2012 JSPS Fellowship for Foreign Researchers.


\end{document}